\documentstyle[epsf,prd,aps,floats]{revtex}

\begin{document}

\draft

\input epsf \renewcommand{\topfraction}{0.8}
\twocolumn[\hsize\textwidth\columnwidth\hsize\csname
@twocolumnfalse\endcsname

\title{
Correlated Isocurvature Fluctuation in Quintessence and
Suppressed CMB Anisotropies at Low Multipoles
}

\author{Takeo Moroi$^{(a)}$ and Tomo Takahashi$^{(b)}$}

\address{
$^{(a)}$Department of Physics, Tohoku University, 
Sendai 980-8578, Japan\\
$^{(b)}$Department of Physics and Astronomy,
University of North Carolina, Chapel Hill, NC 27599
}

\maketitle

\begin{abstract}

We consider cosmic microwave background (CMB) anisotropy in models
with quintessence taking into account of isocurvature fluctuation in
the quintessence.  It is shown that, if the primordial fluctuation of
the quintessence has a correlation with the adiabatic density
fluctuations, CMB angular power spectrum $C_l$ at low multipoles can
be suppressed without affecting $C_l$ at high multipoles.  Possible
scenario of generating correlated mixture of the quintessence and
adiabatic fluctuations is also discussed.

\end{abstract}

\vspace{5mm}

]

\noindent
Preprint number: TU-695, astro-ph/0308208 
\vspace{5mm}

\renewcommand{\thefootnote}{\arabic{footnote}}

\setcounter{footnote}{0}

The recent measurement of the cosmic microwave background (CMB)
angular power spectrum $C_l$ by the Wilkinson Microwave Anisotropy
Probe (WMAP) \cite{aph0302208} has greatly improved our understanding
of the universe.  In particular, since the CMB angular power spectrum
is sensitive to the properties of the origin and evolution of the
cosmic density fluctuations, now we are at the position to test
various scenarios of generating cosmic density fluctuations, which
requires some interplay of astrophysics, cosmology, and particle
physics.  In addition, the shape of the CMB angular power spectrum
also depends on various cosmological parameters, and hence many of the
cosmological parameters are now precisely determined.

One of the important messages from the WMAP is that our universe is
well described by a low-density cold dark matter (CDM) model with
(almost) flat geometry.  In other words, the WMAP confirmed the
existence of the dark energy which is also suggested by the
observations of high red-shift Type-Ia supernovae \cite{SN-Type1}.

Although the cosmological constant is the most famous candidate of the
dark energy, a slowly evolving scalar field, dubbed as
``quintessence'' \cite{Quintessence}, is another important
possibility.  In models with quintessence, however, behavior of the
CMB angular power spectrum may be different from that in the
$\Lambda$CDM model (where the cosmological constant is assumed as the
dark energy).  One reason is that, since the quintessence is a
dynamical scalar field, its amplitude may acquire primordial
fluctuation and that isocurvature fluctuation may exist in the
quintessence sector, which affects behaviors of the CMB angular power
spectrum.

So far, effects of the primordial fluctuation of the quintessence have
been studied only for the cases where the primordial fluctuation of
the quintessence is uncorrelated with the adiabatic density
fluctuations.  In this case, if the quintessence-dominated universe is
realized in very recent epoch as in the $\Lambda$CDM model, effects of
the quintessence fluctuation on $C_l$ are only on very low multipoles.
In addition, in the uncorrelated case, $C_l$ at low multipoles are
always enhanced \cite{PRD64-083009,PLB533-294}.  It is possible,
however, that the primordial fluctuation of the quintessence has
correlation with the adiabatic density fluctuations.  (For explicit
model of generating correlated mixture of the quintessence and
adiabatic fluctuations, see the discussion below.)

Thus, in this letter, we study the CMB angular power spectrum in
models with quintessence paying particular attention to the effects of
the correlation of the quintessence fluctuation with adiabatic
density fluctuations.  If the correlation exists, behavior of the CMB
angular power spectrum becomes different from the uncorrelated case.
In particular, as we will discuss, $C_l$ at low multipoles may be
suppressed without affecting the shape of the CMB angular power
spectrum at high multipoles, which may be related to the suppression
of the measured values of $C_l$ at low multipoles.  (For other models
of suppressing $C_l$ at low multipoles, see \cite{SmallC2}.)

Let us start with presenting the framework.  Here we consider the
scenario where the dark energy of the universe is given by a potential
energy of a slowly evolving scalar field, quintessence $Q$; if the
slow-roll condition is satisfied for the quintessence field $Q$,
equation-of-state parameter of $Q$ is almost $-1$ and the energy
density of $Q$ behaves as the cosmological constant.  Although various
models of quintessence have been proposed, we adopt a simple
approximation of the quintessence potential; we use the parabolic
potential with a constant term:
\begin{eqnarray}
V(Q) = V_0 + \frac{1}{2} m_Q^2 Q^2.
\end{eqnarray}
We assume that $Q$ has non-vanishing initial amplitude.  In our study,
for simplicity, we consider $m_Q$ comparable to (or smaller than) the
present expansion rate of the universe.  In addition, $V_0$ is assumed
to be of the order of the present critical density or smaller.  With
such a small value of $m_Q$ (and $V_0$), slow-roll condition for $Q$
is satisfied until very recently and energy fraction of the
quintessence becomes sizable only at the very recent epoch.  In this
case, shape of the CMB angular power spectrum at high multipoles is
almost unaffected.  Notice that the above potential is a good
approximation for some quintessence models, like the cosine-type one
\cite{JHEP9905-022}.

Since the quintessence is a dynamical scalar field, its amplitude may
fluctuate.  In particular, if its (effective) mass during inflation is
smaller than the expansion rate, the quintessence field also acquires
quantum fluctuation.  Such a primordial fluctuation becomes a new
source of the cosmic density fluctuations and affects the CMB angular
power spectrum.

Evolution and effects of the primordial fluctuation of $Q$ have been
studied for the case where the primordial fluctuation of the
quintessence is not correlated with the adiabatic fluctuations
\cite{PRD64-083513,PRD64-083009,PLB533-294}.  Fluctuation of the
quintessence field can be, however, correlated with the adiabatic
fluctuations.  If some correlation exists, effects of the primordial
fluctuation of $Q$ on the CMB angular power spectrum are expected to
be different from those in the uncorrelated case.  Hereafter, we study
the effects of quintessence fluctuation for the case where the
correlation between the primordial fluctuation of $Q$ and the
adiabatic fluctuations exists.

In order to parameterize the relative size of the primordial
quintessence fluctuation and the adiabatic fluctuations, we define
\begin{eqnarray}
r_Q \equiv 
\frac{\delta Q_{\rm init}}{M_* \Psi_{\rm RD}}.
\end{eqnarray}
(Strictly speaking, the above expression is valid only for the case
where $\delta Q_{\rm init}$ and $\Psi_{\rm RD}$ are fully correlated.
For the case where $\delta Q_{\rm init}$ and $\Psi_{\rm RD}$ are
uncorrelated, for example, it should be understood as
$r_Q=\sqrt{\langle\delta Q_{\rm init}^2\rangle}/M_*
\sqrt{\langle\Psi_{\rm RD}^2\rangle}$.)  Here, $\delta Q_{\rm init}$
is the primordial fluctuation of $Q$, $\Psi$ denotes the fluctuation
of the $(0,0)$ component of the metric in the Newtonian gauge:
$g_{00}=a^2(1+2\Psi)$ with $a$ being the scale factor \cite{PTP78-1},
and $M_*$ is the reduced Planck scale.  In addition, $\Psi_{\rm RD}$
is the metric perturbation related to the adiabatic density
fluctuation in the radiation-dominated epoch.  We assume that
$\Psi_{\rm RD}$ is (almost) scale-invariant, as suggested by the WMAP
\cite{aph0302209}.

If we calculate the CMB angular power spectrum with non-vanishing
values of $\delta Q_{\rm init}$ and $\Psi_{\rm RD}$, we obtain the
form
\begin{eqnarray}
C_l = C_l^{\rm (adi)} + C_l^{\rm (corr)} + C_l^{\rm (uncorr)}.
\end{eqnarray}
Here, $C_l^{\rm (adi)}$ is the result with purely adiabatic density
fluctuations.  With the cosmological and model parameters used in the
following analysis, $C_l^{\rm (adi)}$ almost agrees with the CMB
angular spectrum for the $\Lambda$CDM model $C_l^{\rm (\Lambda CDM)}$.
On the contrary, $C_l^{\rm (uncorr)}$ is the CMB angular power
spectrum purely generated from $\delta Q_{\rm init}$, while $C_l^{\rm
(corr)}$ parameterizes the effects of correlation.

When $\delta Q_{\rm init}$ and $\Psi_{\rm RD}$ are uncorrelated,
$C_l^{\rm (corr)}=0$.  Furthermore, $C_l^{\rm (uncorr)}$ is increased
at low multipoles.  As a result, the total CMB angular power spectrum
may be significantly enhanced at low multipoles
\cite{PRD64-083009,PLB533-294}.

If $\delta Q_{\rm init}$ and $\Psi_{\rm RD}$ have correlation,
$C_l^{\rm (corr)}$ plays important roles.  To see the effects of the
correlation, we calculate the CMB angular power spectrum for the case
where $\delta Q_{\rm init}$ and $\Psi_{\rm RD}$ are fully correlated:
$\langle\delta Q_{\rm init}\Psi_{\rm RD}\rangle^2=\langle\delta Q_{\rm
init}^2\rangle\langle\Psi_{\rm RD}^2\rangle$.  In our numerical
analysis, we always take $Q_{\rm init}>0$.  In this case, primordial
fluctuation of the energy density of $Q$ has positive correlation with
$\Psi_{\rm RD}$ when $r_Q>0$.  (For details of the calculation, see
\cite{PRD64-083009,PLB533-294}.)

\begin{figure}[t]
\begin{center}
\epsfxsize=0.42\textwidth\epsfbox{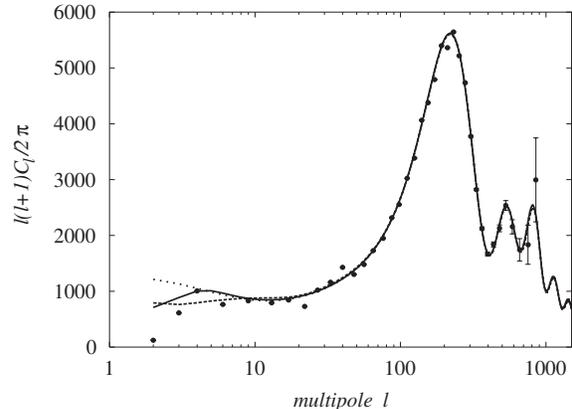}
\caption{ The CMB angular power spectrum generated from the correlated
mixture of the quintessence and adiabatic fluctuations.  We take (a)
$m_Q=10^{-42}\ {\rm GeV}$, $V_0=0$ and $r_Q=400$ with
$\Omega_b=0.046$, $\Omega_m=0.27$, $h=0.72$, and $\tau=0.166$
\protect\cite{aph0302209} (solid), and (b) $m_Q=10^{-40}\ {\rm GeV}$,
$V_0=3.0\times 10^{-47}\ {\rm GeV}$ and $r_Q=2.5$ with
$\Omega_b=0.048$, $\Omega_m=0.27$, $h=0.7$, and $\tau=0.2$ (dashed).
(Here, $\Omega_b$ and $\Omega_m$ are density parameters of baryon and
non-relativistic matter, respectively, $h$ the Hubble constant in
units of 100\ km/sec/Mpc, and $\tau$ the reionization optical depth.)
The full correlation between $\delta Q_{\rm init}$ and $\Psi_{\rm RD}$
is assumed.  Result for the purely adiabatic $\Lambda$CDM model is
also shown in the dotted line.  (For cosmological parameters, we use
the best-fit values suggested by the WMAP for the power-law
$\Lambda$CDM model.)  For comparison, we also plot the data points
measured by the WMAP \protect\cite{aph0302217}.  (The errors are
measurement errors only.)}
\label{fig:Cl}
\end{center}
\end{figure}

In Fig.\ \ref{fig:Cl}, we plot the resultant CMB angular power spectra
for several cases with positive values of $r_Q$: (a) $r_Q=35$ with
$m_Q=10^{-42}\ {\rm GeV}$ and $V_0=0$, and (b) $r_Q=2.5$ with
$m_Q=10^{-40}\ {\rm GeV}$ and $V_0=3.0\times 10^{-47}\ {\rm GeV}$.
(We checked that these data points are consistent with the recent Type
1a supernovae data \cite{Tonry:2003zg}.)  Initial amplitude of the
quintessence $Q_{\rm init}$ is determined so that the present energy
fraction of the quintessence becomes $\Omega_Q=1-\Omega_m=0.73$, which
gives $Q_{\rm init}=8.2 \times 10^{18} $ GeV and $7.5 \times 10^{17}$
GeV for the cases (a) and (b), respectively.  As one can see, if
$r_Q>0$, sizable suppression of the low multipoles is possible
compared to the $\Lambda$CDM model.  Notice that, in the uncorrelated
case, such a suppression of $C_l$ at the low multipoles cannot be
realized.  It is also notable that, even in the correlated case, the
CMB angular power spectrum at higher multipoles is almost the same as
that in the $\Lambda$CDM case.

\begin{figure}[t]
\begin{center}
\epsfxsize=0.42\textwidth\epsfbox{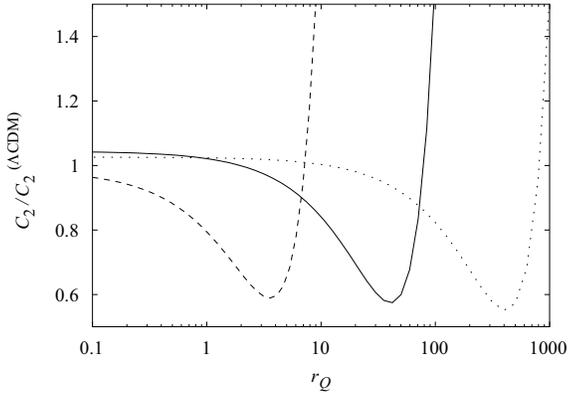}
\caption{ $C_2/C_2^{\rm (\Lambda CDM)}$ as a function of $r_Q$ 
for cases (a) and (b) given in Fig.\ \ref{fig:Cl} (solid and dashed,
respectively), and for $m_Q=10^{-43}\ {\rm GeV}$ and $V_0=0$ (dotted)
with $\Omega_b=0.046$, $\Omega_m=0.27$, $h=0.72$, and $\tau=0.166$.
For $C_l^{\rm (\Lambda CDM)}$, we use the best-fit values of the
cosmological parameters suggested by the WMAP.  The full correlation
between $\delta Q_{\rm init}$ and $\Psi_{\rm RD}$ is assumed.}
\label{fig:ratio}
\end{center}
\end{figure} 

In order to see how much $C_2$ can be suppressed relative to the
$\Lambda$CDM model, in Fig.\ \ref{fig:ratio}, we plot the ratio
$C_2/C_2^{\rm (\Lambda CDM)}$ as a function of $r_Q$.  Since $C_l^{\rm
(corr)}\propto r_Q$ and $C_l^{\rm (uncorr)}\propto r_Q^2$, $C_l^{\rm
(uncorr)}$ is more important than $C_l^{\rm (corr)}$ when $r_Q$ is
large.  In this case, the ratio $C_2/C_2^{\rm (\Lambda CDM)}$ becomes
larger than 1.  If $r_Q$ is smaller, however, $C_l^{\rm (corr)}$
becomes sizable and $C_2$ can be suppressed.  As one can see, with the
correlated primordial fluctuation in the quintessence amplitude,
$C_2/C_2^{\rm (\Lambda CDM)}$ can be as small as $\sim 0.6$ if $r_Q$
is properly chosen.  Importantly, effects of $\delta Q_{\rm init}$ is
limited to $C_l$ with small $l$ and hence the shape of the CMB angular
power spectrum at higher multipoles is unchanged.  Thus, the
correlated fluctuation in the quintessence sector provides a new
mechanism to suppress $C_l$ at very low multipoles without affecting
the high multipoles.  This may have some relevance with the suppressed
values of the CMB angular power spectrum at low multipoles measured by
the WMAP.

Suppression of $C_l$ at low multipoles improves the agreement of the
theoretical prediction with the observation.  To discuss this issue
quantitatively, we calculate the goodness-of-fit parameter $\chi^2$
using the numerical program provided by the WMAP collaboration with
the WMAP data \cite{aph0302217}.  Compared to the $\Lambda$CDM model,
we found that the $\chi^2$ variable can decrease in models with
quintessence.  For example, for the cases (a) and (b) given in Fig.\ 
\ref{fig:Cl}, changes of the goodness-of-fit parameter are
$\Delta\chi^2=-2.1$ and $-1.2$, respectively.  As can be understood
from the figure, the changes in $\chi^2$ are only from the low
multipoles.

In our study, we have also calculated $C_l$ with negative values of
$r_Q$.  Since $C_l^{\rm (corr)}$ is proportional to $r_Q$, the
resultant CMB angular power spectrum is enhanced at the low multipoles
when $r_Q<0$.  Thus, if the suppression of $C_l$ at the low multipoles
measured by the WMAP is related to the primordial fluctuation of the
quintessence, positive correlation with the metric perturbation
$\Psi_{\rm RD}$ (i.e., $r_Q>0$) is required in the present model.

So far, we have discussed effects of primordial fluctuations of
quintessence on the CMB angular power spectrum, so we would like to
comment on other cosmological perturbations in our scenario.  Since
the quintessence dominates the universe at very recent epoch, its
effects are only on fluctuations with very large wavelength comparable
to the present Hubble distance.  In other words, density fluctuations
with smaller wavelength are determined by the adiabatic part of the
fluctuation and hence the results for those perturbations are the same
as the predictions of the simple inflationary paradigm.

Finally, we present a possible scenario which generates the correlated
fluctuations.  We use the fact that fluctuations of two scalar fields
can be correlated if they have a mixing during the inflation
\cite{PRD63-023506}.  We consider the case with two scalar fields, $Q$
and $\phi$.  (In the following, primordial fluctuation of $\phi$
becomes the dominant source of the adiabatic density fluctuations, as
we will see below.)  $Q$ and $\phi$ are defined as mass eigenstates in
the present universe.  In the early universe (i.e., for example,
during inflation), however, Hubble-induced interaction may cause a
mixing between $Q$ and $\phi$ and hence the mass eigenstates may be
linear combinations of them.  We denote the mass eigenstates as $\xi$
and $\eta$, and define the mixing angle $\theta$ as
\begin{eqnarray}
\left( \begin{array}{c}
\xi \\ \eta
\end{array} \right) = 
\left( \begin{array}{cc}
\cos\theta (t) & -\sin\theta (t) \\ \sin\theta (t) & \cos\theta (t)
\end{array} \right)
\left( \begin{array}{c}
Q \\ \phi
\end{array} \right).
\end{eqnarray}
In our model, $\theta (t)$ varies from non-vanishing value during
inflation $\theta_{\rm inf}$ to the present value $\theta_{\rm
now}=0$.  If the (effective) mass of $\eta$ is large during inflation
while that of $\xi$ is negligible, then only the $\xi$ field acquires
the quantum fluctuation:
\begin{eqnarray}
\delta \xi_{\rm inf} = \frac{H_{\rm inf}}{2\pi},~~~
\delta \eta_{\rm inf} = 0,
\end{eqnarray}
where $H_{\rm inf}$ is the expansion rate during the inflation.
Assuming that $\theta (t)$ rapidly changes from $\theta_{\rm inf}$ to
$0$ at the end of inflation, primordial fluctuation of $Q$ and $\phi$
are given by
\begin{eqnarray}
\delta Q_{\rm init} = \delta \xi_{\rm inf}\cos\theta_{\rm inf},~~~
\delta \phi_{\rm init} = \delta \xi_{\rm inf} \sin\theta_{\rm inf},
\label{dQ&dphi}
\end{eqnarray}
and correlated fluctuations are generated in the quintessence and
$\phi$ fields.

The above situation may be realized if the potential of the scalar
fields is of the form
\begin{eqnarray}
V = V_0 + \frac{1}{2} m_Q^2 Q^2 + \frac{1}{2} m_\phi^2 \phi^2 
+ V_{\rm Hubble},
\label{V_tot}
\end{eqnarray}
where
\begin{eqnarray}
V_{\rm Hubble} = H_{\rm vac}^2 
( Q \sin \theta_{\rm inf} + \phi \cos \theta_{\rm inf} )^2.
\label{V_H}
\end{eqnarray}
Here, $V_{\rm Hubble}$ is the Hubble-induced interaction which is
effective only during the inflation with $H_{\rm vac}$ being the
expansion rate of the universe induced by the ``vacuum energy;''
$H_{\rm vac}=H_{\rm inf}$ and $H_{\rm vac}\simeq 0$ for during and
after the inflation, respectively.  With this potential, one of the
mass eigenstates $\eta\simeq\phi+\theta_{\rm inf}Q$ acquires an
effective mass comparable to the expansion rate and its quantum
fluctuation during inflation becomes negligibly small.  Other mass
eigenstate $\xi\simeq Q-\theta_{\rm inf}\phi$, on the contrary, stays
almost massless and it acquires the quantum fluctuation.

If the decay rate of the inflaton field is larger than $m_\phi$, slow
roll condition is satisfied for $\phi$ at the time of the inflaton
decay.  In this case, $\delta\phi_{\rm init}$ may become the dominant
source of the adiabatic density fluctuations.  In order to generate
the adiabatic fluctuations from the fluctuation of $\phi$, we can use
the curvaton mechanism \cite{curvaton} where the primordial
fluctuation of the curvaton, a late-decaying scalar field, becomes the
dominant source of the adiabatic density fluctuations.  (Here, we do
not identify $\phi$ as inflaton; in our model, $\phi$ should acquire
large effective mass during inflation and hence $\phi$ cannot be the
inflaton.)  Indeed, if the energy density of $\phi$ once dominates the
universe, $\phi$ plays the role of curvaton and the metric
perturbation in the radiation dominated epoch is given by
\cite{Psi(curv)}
\begin{eqnarray}
\Psi_{\rm RD} = -\frac{4}{9} 
\frac{\delta\phi_{\rm init}}{\phi_{\rm init}},
\end{eqnarray}
where $\phi_{\rm init}$ is the initial amplitude of $\phi$ determined
during the inflation.  As a result, with Eq.\ (\ref{dQ&dphi}),
correlated mixture of adiabatic and quintessence fluctuations is
generated.  In this model, the $r_Q$ parameter is estimated as
\begin{eqnarray}
r_Q = \frac{1}{2\pi} \frac{H_{\rm inf}}{M_* \Psi_{\rm RD}}
\cos\theta_{\rm inf}.
\end{eqnarray}
Notice that, using $\Psi_{\rm RD}\sim O(10^{-5})$ and the upper bound
$H_{\rm inf}/M_*\lesssim 7\times 10^{-5}$ \cite{H_max}, $r_Q\lesssim
1$ in this simple model.  Larger value of $r_Q$ is, however, possible
if we extend the model.  For example, if the coefficient of the
kinetic term of $Q$ varies after the inflation, value of $\delta Q$
(for the canonically normalized field) also changes.  Including this
effect,
\begin{eqnarray}
r_Q = Z^{-1/2}_{Q, {\rm inf}}
\frac{H_{\rm inf}/2\pi}{M_* \Psi_{\rm RD}}
\cos\theta_{\rm inf},
\end{eqnarray}
where $Z_{Q, {\rm inf}}$ is the coefficient of the kinetic term of $Q$
during the inflation.  (We normalize the present value of $Z_{Q}$ to
be $1$.)  Thus, if $Z_{Q, {\rm inf}}<1$, $r_Q$ can be enhanced.

In summary, we have seen that the CMB angular power spectrum at small
$l$ can be suppressed without affecting that at higher multipoles in
models with quintessence, if the primordial fluctuation of the
quintessence has correlation with the adiabatic density fluctuations.
We have also pointed out that such a correlation may be generated
during inflation if the quintessence field has some mixing with other
scalar field which is responsible for generating the adiabatic density
fluctuations.

{\sl Acknowledgments:} We acknowledge the use of CMBFAST
\cite{cmbfast} package for our numerical calculations.  T.T. thanks
High Energy Theory Group in Tohoku University, where this work has
been done, for their hospitality during the visit.  The work of
T.M. is supported by the Grant-in-Aid for Scientific Research from the
Ministry of Education, Science, Sports, and Culture of Japan, No.\ 
15540247.  The work of T.T. is supported by the US Department of
Energy under Grant No. DE-FG02-97ER-41036.


\begin{references}
\vspace{-12mm}

\bibitem{aph0302208}
    C.~Bennett {\it et al.},
    arXiv:astro-ph/0302208.

\bibitem{SN-Type1}
    A.~G.~Riess {\it et al.},
    Astron.\ J.\  {\bf 116} (1998) 1009;
    S.~Perlmutter {\it et al.},
    Astrophys.\ J.\  {\bf 517} (1999) 565.

\bibitem{Quintessence}
    P.~J.~Peebles and B.~Ratra,
    Astrophys.\ J.\  {\bf 325} (1988) L17;
    B.~Ratra and P.~J.~Peebles,
    Phys.\ Rev.\ D {\bf 37} (1988) 3406;
    R.~R.~Caldwell, R.~Dave and P.~J.~Steinhardt,
    Phys.\ Rev.\ Lett.\  {\bf 80} (1998) 1582;
    S.~M.~Carroll,
    Phys.\ Rev.\ Lett.\  {\bf 81} (1998) 3067;
    I.~Zlatev, L.~M.~Wang and P.~J.~Steinhardt,
    Phys.\ Rev.\ Lett.\  {\bf 82} (1999) 896.

\bibitem{PRD64-083009}
    M.~Kawasaki, T.~Moroi and T.~Takahashi,
    Phys.\ Rev.\ D {\bf 64} (2001) 083009.

\bibitem{PLB533-294}
    M.~Kawasaki, T.~Moroi and T.~Takahashi,
    Phys.\ Lett.\ B {\bf 533} (2002) 294.

\bibitem{SmallC2}
    C.~R.~Contaldi, M.~Peloso, L.~Kofman and A.~Linde,
    JCAP {\bf 0307} (2003) 002;
    J.~M.~Cline, P.~Crotty and J.~Lesgourgues,
    arXiv:astro-ph/0304558;
    B.~Feng and X.~Zhang,
    arXiv:astro-ph/0305020;
    M.~Kawasaki and F.~Takahashi,
    arXiv:hep-ph/0305319;
    M.~Bastero-Gil, K.~Freese and L.~Mersini-Houghton,
    arXiv:hep-ph/0306289;
    N.~Kaloper and M.~Kaplinghat,
    arXiv:hep-th/0307016;
    S.~Tsujikawa, R.~Maartens and R.~Brandenberger,
    arXiv:astro-ph/0308169.

\bibitem{JHEP9905-022}
    J.~E.~Kim,
    JHEP {\bf 9905} (1999) 022.

\bibitem{PRD64-083513}
    L.~R.~Abramo and F.~Finelli,
    Phys.\ Rev.\ D {\bf 64} (2001) 083513.

\bibitem{PTP78-1}
    H.~Kodama and M.~Sasaki,
    Prog.\ Theor.\ Phys.\ Suppl.\  {\bf 78} (1984) 1.

\bibitem{aph0302209}
    D.~N.~Spergel {\it et al.},
    arXiv:astro-ph/0302209.


\bibitem{Tonry:2003zg}
    J.~L.~Tonry {\it et al.},
    Astrophys.\ J.\  {\bf 594} (2003) 1

\bibitem{aph0302217}
    G.~Hinshaw {\it et al.},
    arXiv:astro-ph/0302217.

\bibitem{PRD63-023506}
    C.~Gordon, D.~Wands, B.~A.~Bassett and R.~Maartens,
    Phys.\ Rev.\ D {\bf 63} (2001) 023506.

\bibitem{curvaton}
    S.~Mollerach,
    Phys.\ Rev.\ D {\bf 42} (1990) 313;
    A.~D.~Linde and V.~Mukhanov,
    Phys.\ Rev.\ D {\bf 56} (1997) 535;
    K.~Enqvist and M.~S.~Sloth,
    Nucl.\ Phys.\ B {\bf 626} (2002) 395;
    D.~H.~Lyth and D.~Wands,
    Phys.\ Lett.\ B {\bf 524} (2002) 5;
    T.~Moroi and T.~Takahashi,
    Phys.\ Lett.\ B {\bf 522} (2001) 215
    [Erratum-ibid.\ B {\bf 539} (2002) 303].
    
\bibitem{Psi(curv)}
    T.~Moroi and T.~Takahashi, in Ref.\ \cite{curvaton};
    Phys.\ Rev.\ D {\bf 66} (2002) 063501.

\bibitem{H_max}
    H.~V.~Peiris {\it et al.},
    Astrophys.\ J.\ Suppl.\  {\bf 148} (2003) 213;
    V.~Barger, H.~S.~Lee and D.~Marfatia,
    Phys.\ Lett.\ B {\bf 565} (2003) 33;
    S.~M.~Leach and A.~R.~Liddle,
    arXiv:astro-ph/0306305.
    
\bibitem{cmbfast}
    U.~Seljak and M.~Zaldarriaga,
    Astrophys.\ J.\  {\bf 469} (1996) 437.
    
\end{references}
\end{document}